# Topology optimization of micro-structured materials featured with the specific mechanical properties

Jie Gao[1,2], Hao Li[1], Zhen Luo[2], Liang Gao[1*], Peigen Li[1]

[1] *The State Key Laboratory of Digital Manufacturing Equipment and Technology, Huazhong University of Science and Technology, 1037 Luoyu Road, Wuhan, Hubei 430074, China*

[2] *School of Mechanical and Mechatronic Engineering, University of Technology Sydney, 15 Broadway, Ultimo, NSW 2007, Australia*

Telephone number of the corresponding author: +86-27-87557742;

Fax of the corresponding author: +86-27-87559419

**Abstract:** Micro-structured materials consisting of an array of microstructures are engineered to provide the specific material properties. This present work investigates the design of cellular materials under the framework of level set, so as to optimize the topologies and shapes of these porous material microstructures. Firstly, the energy-based homogenization method (EBHM) is applied to evaluate the material effective properties based on the topology of the material cell, where the effective elasticity property is evaluated by the average stress and strain theorems. Secondly, a parametric level set method (PLSM) is employed to optimize the microstructural topology until the specific mechanical properties can be achieved, including the maximum bulk modulus, the maximum shear modulus and their combinations, as well as the negative Poisson's ratio (NPR). The complicated topological shape optimization of the material microstructure has been equivalent to evolve the sizes of the expansion coefficients in the interpolation of the level set function. Finally, several numerical examples are fully discussed to demonstrate the effectiveness of the developed method. A series of new and interesting material cells with the specific mechanical properties can be found.

# 1 Introduction

Micro-structured materials consisting of a number of arranged periodical material microstructures are characterized with high performance but low mass [Ashby et al. (2000); Gibson and Ashby (1999)], which have gradually gained considerable applications in engineering, such as energy absorption, thermal insulation and etc. It is known that the extraordinary properties of micro-structured materials mainly depend on the topologies of material cells rather than the constitutive compositions [Gibson and Ashby (1999)]. Although an increasing number of publications have been reported to improve material properties by modifying the geometrical dimensions [Christensen (2000)], a systematic and effective design method for micro-structured materials is still in demand.

Topology optimization has long been recognized as a powerful tool to find the optimal design of both structures [Sigmund and Maute (2013)] and materials [Cadman et al. (2012); Osanov and Guest (2016)] in recent years, which has spread into a wide variety of applications, like the dynamic [Ma et al. (1993)], the engineering structure "blade" [DR McClanahan et al. (2018)], the stress problem [Chu et al. (2018); Xiao et al. (2018)]; the multi-functional materials [Cadman et al. (2012); Radman et al. (2012)] and the concurrent design [Groen et al. (2018); Rodrigues et al. (2002); Wang et al. (2017); Wang et al. (2018); Wang et al. (2018); Xia et al (2015)]. The basic idea is that materials are iteratively eliminated and redistributed in a defined design space to seek the best structural topologies with the optimal performance. Now, many topology optimization methods are proposed, like homogenization method [Bendsøe and Kikuchi (1988); Guedes and Kikuchi (1990)], the solid isotropic material with penalization (SIMP) method [Bendsøe and Sigmund (1999); Zhou and Rozvany (1991)], evolutionary structural optimization [Xie and Steven (1993)] and the level set method (LSM) [Allaire et al. (2004); Sethian and Wiegmann (2000); Wang et al. (2003)]. Compared with the material distribution models [Bendsøe and Kikuchi (1988); Gao et al. (2017); Kang et al. (2011); Radman et al. (2013)], the LSM opitimzes the structural topologies by evolving the structural boundaries rather than the densities, so that the optimized structures are characterized with smooth boundaries and clear interfaces which would be directly integrated into the system of CAD/CAM. In fact, the LSM has several merits in terms of the geometrical and physical views, like (1) smooth and distinct boundary description, (2) shape fidelity and higher topological flexibility, (3) shape and topology optimization simultaneously and (4) a physical meaning solution of the Hamilton-Jacobi partial differential equation (H-J PDE).

The LSM [Osher and Sethian (1988)] has been first introduced into the field of structural optimization by [Sethian and Wiegmann (2000)]. The concept of the LSM is to implicitly represent the structural boundary as the zero-level set of a higher-dimensional scalar function, namely the level set function. Then, the evolvement of the level set front is tracked and driven by the mathematical solving of first-order H-J PDE [Allaire et al. (2004); Wang et al. (2003)]. After that, it has been quickly expanded to solve a broad range of topological shape optimization problems, like the dynamic [Li et al. (2017)], multi-phase [Wang and Wang (2004); Kang et al. (2016)], the manufacturing constraint [Wang et al. (2018); Zhang et al (2018)], structure-material integrated design [Wang et al. (2017)] and compliant mechanism [Wang et al. (2005)]. It should be noted that the PDE-driven LSM can be classified as the standard or conventional LSM. There are also several unfavourable features of the standard LSM aroused from the direct solving of the H-J PDE by the up-wind schemes, e.g. the re-initializations, extension of the velocity field, and CFL condition [Allaire et al. (2004); Wang et al. (2003)].

In response to the numerical difficulties in the standard LSM, several variants of LSMs have been proposed, such as introducing the hole insertion mechanism [Dunning and Alicia Kim (2013), explicit level set [Guo et al. (2014)], combing Semi-Lagrange method [Xia et al. (2006)] and the interpolation by radial basis functions (RBFs) [Wang and Wang (2006)]. One of variants termed by the parametric level set method (PLSM) [Luo et al. (2009); Luo et al. (2008)] has been considered as a powerful alternative level set method, which can not only inherit the positive merits, but also eliminate the unfavorable features of the standard LSM. In the PLSM, the level set function is interpolated by the CSRBFs, so that the advancing of structural boundaries is equivalent to the evolution of the sizes of expansion coefficients. Hence, the complicated structural optimization has been transformed into a much easier "size" optimization problem. Moreover, many well-established efficient optimization algorithms can be directly applied into the PLSM to improve the optimization efficiency, e.g. the optimality criteria (OC) [Rozvany et al. (1996)] and method of moving asymptotes (MMA) [Svanberg (1987)]. Many optimization problems are also solved perfectly by the PLSM, like the manufacturing constraint [Li et al. (2015)], functional graded materials [Li et al. (2018); Li et al. (2016)], multi-materials [Wang et al. (2015)] and robust optimization [Wu et al. (2016)].

Since the homogenization [Guedes and Kikuchi (1990)] has been established, it is rapidly becoming popular to combine with topology optimization to develop an inverse design procedure for seeking the best topologies of PUCs with superior material effective properties [Guest and Prévost (2006); Huang et al. (2011); Long et al. (2016); Sigmund (1994); Torquato et al. (2003)]. The homogenization is utilized to evaluate material effective properties which work as the objective function, and topology optimization is applied to evolve the topology of the material cell until the expected effective property is gained. After the pioneering work of Sigmund [Sigmund (1994)], various materials with novel physical performances have been presented, such as extreme thermal properties, maximum stiffness and fluid permeability [Challis et al. (2012); Guest and Prévost (2006)] and negative Poisson's ratio [Wang et al. (2014)]. Hence, the design of micro-structured materials has become one of the most promising applications associated with topology optimization [Berger et al. (2017); Cadman et al. (2012); Cadman et al. (2013); Osanov and Guest (2016)]. Nevertheless, the majority of these existing works are based on the material density-related interpolation schemes, while the positive features of the LSM are beneficial to the optimization of micro-structured materials, particularly considering the manufacturing.

In majority of the mentioned-above works, the numerical homogenization method (NHM) is used to evaluate material effective properties [Berger et al. (2017); Cadman et al. (2013); Li Eric et al. (2015); Li Eric et al. (2016); Osanov and Guest (2016)]. It is known that the asymptotic expansion theory is the basic theoretical framework of the NHM [Hassani and Hinton (1998a); Hassani and Hinton (1998b)], while theoretical derivations and numerical implementations are relatively complex. For example, a fictitious body force needs to be formulated by the imposing of initial unit test strains to solve the induced strains. As an alternative way, the energy-based homogenization method (EBHM) [Sigmund (1994); Xia and Breitkopf (2015)] has been successfully developed, where the average stress and strain theorems are the main theoretical basis to predict the effective properties. Compared with the NHM, the periodic boundary conditions are imposed on the structural boundaries of PUCs to transform the initial elastic equilibrium equation of the finite element analysis into a much more compact form [Gao et al. (2018); Xia et al. (2016); Da et al. (2017a)]. Hence, the numerical solution of the EBHM to predict effective properties is much easier. Moreover, the simple theoretical analysis of the EBHM is beneficial to directly connect with topology optimization.

In this paper, we develop a new topology optimization method for the design of the micro-structured materials by the PLSM combined with the EBHM to attain the specific mechanical properties, which includes the maximum bulk modulus, the maximum shear modulus and their combination, as well as the NPR. The remainder is organized as follows. A detailed introduction about the PLSM is provided in Section 2 and the EBHM is in Section 3. The optimization formulation for the micro-structured materials design is developed in Section 4. The numerical implementations are presented in Section 5, and several numerical examples are provided to display the optimized material microstructures in Section 6. Finally, conclusions are drawn in Section 7.

## 2 Parametric level set method

### 2.1 Level set-based boundary representation

The idea of the LSM is that the structural design boundary is implicitly described by the zero-level set of the level set function with a higher dimension, shown in **Fig. 1**. A 2D design domain can be represented by the level set function, as follows:

$$\begin{cases} \Phi(\mathbf{x}) > 0, & \forall \mathbf{x} \in \Omega \setminus \Gamma \quad (material) \\ \Phi(\mathbf{x}) = 0, & \forall \mathbf{x} \in \Gamma \cap \quad \quad \quad ndary) \\ \Phi(\mathbf{x}) < 0, & \forall \mathbf{x} \in D \setminus \Omega \quad (void) \end{cases} \quad (1)$$

where $\Omega$ is the structural design domain contained in reference domain $D$. The term $\Gamma$ denotes the structural boundary at the zero-level set.

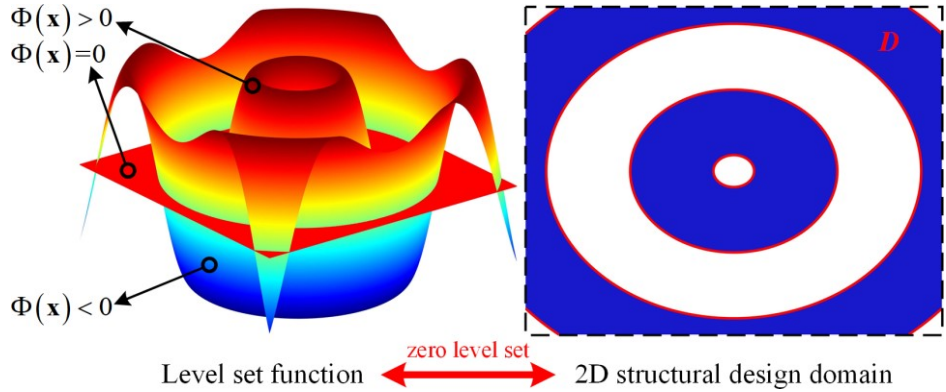

**Fig. 1**. 3D LSF and 2D structural design domain

Introducing a pseudo-time $t$ into the LSM and differentiating on both sides with respect to time variable, we can gain a Hamilton-Jacobi partial differential equation (H-J PDE) which derive the dynamic evolution of the structural boundary [Osher and Sethian (1988); Sethian and Wiegmann (2000)], so that the merging and splitting of the boundaries towards its optimal can be mathematically governed by iteratively solving of the H-J PDE.

$$\frac{\partial \Phi(\mathbf{x},t)}{\partial t} - \upsilon_n \left|\nabla \Phi(\mathbf{x},t)\right| = 0 \tag{2}$$

where $\upsilon_n$ is the normal velocity field at the structural boundary. Structural topology optimization by using the LSM can be regarded as a dynamic evolution process of the level set surface via updating the normal velocity $\upsilon_n$. However, as mentioned previously, direct solving of the H-J PDE to obtain the velocity field along boundary requires the complicated numerical schemes [Allaire et al. (2004); Wang et al. (2003)], which is known as the main disadvantage of the standard LSM.

*2.2 CSRBF-based level set parameterization*

In order to response to these numerical shortcomings, the RBFs [Buhmann (2003)] is utilized to interpolate the level set function. Although the global supported radial basis functions (GSRBF) show the superior performances in some aspects, the full dense matrix and the selection of free shape parameter $c_i$ severely affect the efficiency of the GSRBF [Wang and Wang (2006)]. Compared to the GSRBF, the compactly supported RBFs (CSRBF) with matrix sparseness and strictly positive definiteness can easily inherit the Lipschitz continuity of the interpolation [Wendland (1995)], which show the great potentials in dealing with the topological shape optimization problems. In this work, the CSRBF with C2 continuity presented by Wendland [Wendland (1995)] is adopted:

$$\phi(r) = (1-r)_+^4 \cdot (4r+1) \quad (\textit{Wendland-C2}) \tag{3}$$

where *r* is the radius of support defined in a two-dimensional Euclidean space, as:

$$r = \frac{d_I}{d_{mI}} = \frac{\sqrt{(x-x_i)^2 + (y-y_i)^2}}{d_{mI}} \tag{4}$$

where $d_I$ denotes the Euclidean distance between the current sample knot (*x, y*) and the knot ($x_i$, $y_i$) within whole design domain. The parameter $d_{mI}$ defines the size of the influence domain of the sample knot (*x, y*). The dynamic LSF can be approximated by the product of a series of CSRBFs and their expansion coefficients at different knots [Luo et al. (2009); Luo et al. (2008)]:

$$\Phi(\mathbf{x},t) = \boldsymbol{\varphi}(\mathbf{x})^T \boldsymbol{\alpha}(t) = \sum_{i=1}^{N} \phi_i(x)\alpha_i(t) \tag{5}$$

The series of CSRBFs can be formulated as a vector at every knot:

$$\boldsymbol{\varphi}(\mathbf{x}) = \begin{bmatrix} \phi_1(\mathbf{x}) & \phi_2(\mathbf{x}) & \cdots & \end{bmatrix}^T \tag{6}$$

The series of expansion coefficients is then expressed as:

$$\boldsymbol{\alpha}(t) = \begin{bmatrix} \alpha_1(t) & \alpha_2(t) & \cdots & \end{bmatrix}^T \tag{7}$$

In Eq. (7) the CSRBF term $\boldsymbol{\varphi}(\mathbf{x})$ is only spatial-dependent and the expansion coefficient vector $\boldsymbol{\alpha}$ is only time-dependent. Substituting Eq. (5) into Eq. (2), the initial H-J PDE can be rewritten as:

$$\boldsymbol{\varphi}(\mathbf{x})^T \frac{d\boldsymbol{\alpha}(t)}{dt} - \upsilon_n \left| (\nabla \boldsymbol{\varphi}(\mathbf{x}))^T \boldsymbol{\alpha}(t) \right| = 0 \tag{8}$$

The normal velocity $\upsilon_n$ can be given by:

$$\upsilon_n = \frac{\boldsymbol{\varphi}(\mathbf{x})^T}{\left| (\nabla \boldsymbol{\varphi}(\mathbf{x}))^T \boldsymbol{\alpha}(t) \right|} \frac{d\boldsymbol{\alpha}(t)}{dt} \tag{9}$$

Hereto, the standard LSM is transformed into a parametric form. It can be easily seen that $\upsilon_n$ is naturally extended to the whole design domain, and there is no need to employ additional schemes to extend the velocity field. It can be easily seen that the initial H-J PDE is transformed into a series of ordinary differential equations (ODEs) only with the unknown expansion coefficients. The solving of the complicated H-J PDE to derive the evolution of the structural topology has been converted into evolve the sizes of the expansion coefficients in ODEs, namely a relatively easier "size" optimization problem and the expansion coefficients acting as the variables [Li et al. (2015); Luo et al. (2008)].

## 3 Energy-based homogenization method

The homogenization is utilized to predict the homogenized effective properties of micro-structured materials by analysing the information at the microscale [Hassani and Hinton (1998a)]. In the scope of linear elasticity, the local coordinate system $\mathbf{y}$ ($[0 \ y_1] \times [0 \ y_2]$) is utilized to define material cell in the global coordinate system $\mathbf{x}$ ($[0 \ x_1] \times [0 \ x_2]$). The elastic property $E^\xi(\mathbf{x})$ is $\mathbf{y}$-periodic function in the coordinate system $\mathbf{x}$. In this case, the displacement field inside PUCs can be characterized by the asymptotic expansion theory. For numerical simplicity, only the first-order term of the small parameter expansion is considered. In this case, the homogenized elasticity tensor can be calculated according to material distributions in the material cell.

$$E^H_{ijkl} = \frac{1}{|Y|} \int_\Omega \left( \varepsilon_{pq}^{0(ij)} - \varepsilon_{pq}^{*(ij)} \right) E_{pqrs} \left( \varepsilon_{rs}^{0(kl)} - \varepsilon_{rs}^{*(kl)} \right) d\Omega \quad (i,j,k,l = 1,2,\cdots \tag{10}$$

where $|Y|$ denotes the area in 2D or the volume in 3D of the material cell, and $\varepsilon_{pq}^{0(ij)}$ is the initial unit strain. The unknown strain field $\varepsilon_{pq}^{*(ij)}$ is solved by the following linear elasticity equilibrium equation with **y**-periodicity [Hassani and Hinton (1998a)]:

$$\int_\Omega E_{ijpq}\varepsilon_{rs}^{*(kl)}\frac{\partial v_i}{\partial y_j}d\Omega = \int_\Omega E_{ijpq}\varepsilon_{rs}^{0(kl)}\frac{\partial v_i}{\partial y_j}d\Omega \quad \forall v_i \in H_{per}(\Omega,\mathbf{R}^3) \tag{11}$$

where the term $v_i$ is virtual displacement fields belonging to the **y**-periodic Sobolev functional space. In the finite element analysis, the material cell is assumed to use a mesh of $N_E$ finite elements. In this way, the homogenized elasticity tensor $\mathbf{E}^H$ is reformulated as the sum of the integration over finite elements, given by:

$$E_{ijkl}^H = \frac{1}{|Y|}\sum_{e=1}^{N_E}\left(\mathbf{u}_e^{0(ij)}-\mathbf{u}_e^{*(ij)}\right)^T \mathbf{k}_e \left(\mathbf{u}_e^{0(kl)}-\mathbf{u}_e^{*(kl)}\right) \tag{12}$$

where $\mathbf{u}_e^{*(ij)}$ is the unknown element displacements, and $\mathbf{k}_e$ is the element stiffness matrix. In the EBHM, the initial unit test train is directly imposed on the boundaries of PUCs. The induced strain field in PUCs corresponds to the superimposed strain fields $\left(\varepsilon_{pq}^{0(ij)}-\varepsilon_{pq}^{*(ij)}\right)$ in Eq. (10), which is denoted by the symbol $\varepsilon_{pq}^{Id(ij)}$. Eq. (12) is transformed into a new form in terms of the elementary mutual energies [Sigmund (1994)] by the induced displacement field in PUCs, given as:

$$E_{ijkl}^H = \frac{1}{|Y|}\sum_{e=1}^{N_E}Q_{ijkl}^e = \frac{1}{|Y|}\sum_{e=1}^{N_E}\left(\mathbf{u}_e^{Id(ij)}\right)^T \mathbf{k}_e \mathbf{u}_e^{Id(kl)} \tag{13}$$

where $\mathbf{u}_e^{Id}$ is the corresponding induced element displacements, and $Q_{ijkl}^e$ stands for the elementary mutual energy. The effective elasticity properties are interpreted as the summation of elastic energies of PUCs [Michel et al. (1999); Sigmund (1994)].

The EBHM and the NHM can be regarded as two typical numerical methods for the homogenization to evaluate the homogenized elasticity properties. There exist several differences among them: (1) The asymptotic expansion theory is the basis of the NHM, and the average stress and strain theorems are the key of the EBHM; (2) A uniform periodical boundary condition is constructed to compute the displacement field in the NHM, but the simplified periodic boundary conditions are imposed in the EBHM; (3) The linearly elastic equilibrium equation is reduced in the EBHM based on the definition of the periodic boundary conditions. Hence, the EBHM will be beneficial to improve the computing efficiency of material effective properties [Gao et al. (2018); Xia and Breitkopf (2015)].

## 4 Optimization of micro-structured materials

### *4.1 Optimization model*

The basic idea of the micro-structured materials design is clearly displayed in **Fig. 2**. The intention is to obtain the material cell with the specific mechanical property, which includes the maximum bulk modulus, maximum shear modulus and their combination as well as the NPR.

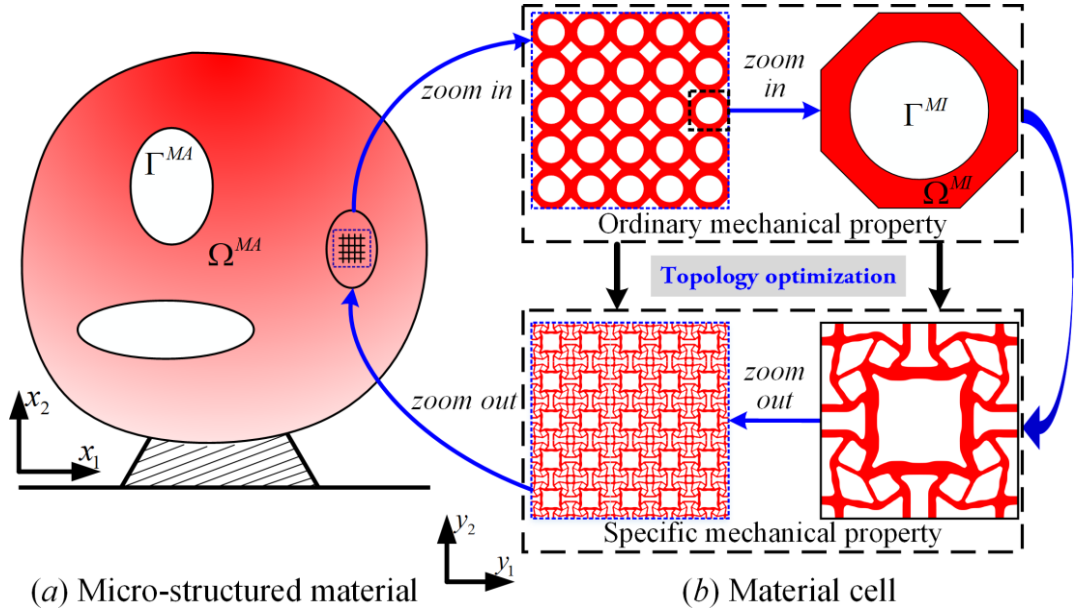

(*a*) Micro-structured material  (*b*) Material cell
**Fig. 2.** The optimization of micro-structured materials

The optimization model is established by integrating the PLSM with the EBHM under the allowable material consumption, given as:

$$\begin{aligned}
&Find: \quad \boldsymbol{\alpha}\left(\alpha_1, \alpha_2, \cdots \cdots\right) \\
&\underset{(\mathbf{u}^*, \Phi)}{Maximize}: J\left(\mathbf{u}^*, \Phi\right) = \sum_{i,j,k,l=1}^{d} \eta_{ijkl} E_{ijkl}^{H}(\boldsymbol{\alpha}) \\
&Subject\ to: \begin{cases} V(\Phi) = \int_{\Omega} H(\Phi) d\Omega - V_{max} \leq 0 \\ a(\mathbf{u}^*, \mathbf{v}, \Phi) = l(\mathbf{v}, \Phi) \quad \forall \mathbf{v} \in H_{per}(\Omega, \mathbf{R}^3) \\ \bar{\alpha}^L \leq \bar{\alpha}_i \leq \bar{\alpha}^U \quad (i=1,2,\cdots \end{cases}
\end{aligned} \quad (14)$$

where $\boldsymbol{\alpha}$ denotes the vector of design variables. $\bar{\alpha}^L$ and $\bar{\alpha}^U$ are the lower and upper bounds of the regularized design variables $\bar{\alpha}_i$, respectively. The initial values of the expansion coefficients are evaluated by the initial level set function, shown in Eq. (5). The *J* is objective function defined by a proper combination of the homogenized elasticity tensor $E_{ijkl}^H$ and the weight factor $\eta_{ijkl}$. $V$ is the volume fraction constraint with an upper bound of $V_{max}$. $H$ is the Heaviside function which serves as a characteristic function (Wang et al. 2003), given as:

$$H(\Phi) = \begin{cases} \eta, & \Phi < -\Delta \\ \dfrac{3(1-\eta)}{4}\left(\dfrac{\Phi}{\Delta} - \dfrac{\Phi^3}{3\Delta^3}\right) + \dfrac{1+\eta}{2}, & -\Delta \leq \Phi \leq \Delta \\ 1, & \Phi > \Delta \end{cases} \quad (15)$$

where $\eta$ is a small positive number to avoid singularity of the numerical process, and $\Delta$ describes the width for numerical approximation of $H$. The homogenized effective tensor $E_{ijkl}^H$ is evaluated by the EBHM in Eq. (13). The linear elasticity equilibrium equation for the material cell is stated in the weak variational form, in which the bilinear energy term $a$ and the linear load form $l$ are written as:

$$\begin{cases} a(\mathbf{u}^*, \mathbf{v}, \Phi) = \int_D \varepsilon_{ij}(\mathbf{u}^*) E_{ijkl} \varepsilon_{kl}(\mathbf{v}) H(\Phi) d\Omega \\ l(\mathbf{v}, \Phi) = \int_D \varepsilon_{ij}(\mathbf{u}^0) E_{ijkl} \varepsilon_{kl}(\mathbf{v}) H(\Phi) d\Omega \end{cases} \quad (16)$$

In the optimization model defined in Eq. (14), the expansion coefficients are identified as the design variables and updated by the much more efficient gradient-based optimization algorithms, rather than directly solving the H-J PDE using the up-wind scheme.

*4.2 Sensitivity analysis*

In the gradient-based optimization algorithms [Svanberg (1987); Zhou and Rozvany (1991)], the first-order derivatives of the objective and constraint functions with respect to the design variables are required. The first-order derivatives of the objective associated with the design variables is given as:

$$\frac{\partial J}{\partial \boldsymbol{\alpha}} = \sum_{i,j,k,l=1}^{d} \left(\eta_{ijkl} \frac{\partial E_{ijkl}^H}{\partial \boldsymbol{\alpha}}\right) \quad \boldsymbol{\alpha} = [\alpha_1 \quad \alpha_2 \quad \cdots] \quad (17)$$

The first-order derivative of the homogenized elastic tensor with respect to time $t$ can be given by:

$$\frac{dE_{ijkl}^H}{dt} = \frac{1}{|\mathbf{Y}|} \int_D \left(\varepsilon_{pq}^{0(ij)} - \varepsilon_{pq}^{*(ij)}\right) E_{pqrs} \left(\varepsilon_{rs}^{0(kl)} - \varepsilon_{rs}^{*(kl)}\right) \upsilon_\mathbf{n} |\nabla \Phi| \delta(\Phi) d\Omega \quad (18)$$

where $\delta(\Phi)$ is the first-order derivative of the Heaviside function with respect to time, namely Dirac function (Osher and Fedkiw 2006). Substituting the velocity defined in Eq. (9) into Eq. (18), it yields:

$$\frac{dE_{ijkl}^{H}}{dt} = \frac{1}{|Y|} \int_{D} \left( \varepsilon_{pq}^{0(ij)} - \varepsilon_{pq}^{*(ij)} \right) E_{pqrs} \left( \varepsilon_{rs}^{0(kl)} - \varepsilon_{rs}^{*(kl)} \right) \left( \boldsymbol{\varphi}(\mathbf{x})^{T} \frac{d\boldsymbol{\alpha}(t)}{dt} \right) \delta(\Phi) d\Omega \tag{19}$$

Since the LSF has been decoupled in space and time by the CSRBF interpolation, the expansion coefficient vector $\boldsymbol{\alpha}$ only depends on time variable, and the above form is rewritten as:

$$\frac{dE_{ijkl}^{H}}{dt} = \frac{1}{|Y|} \left\{ \int_{D} \left( \varepsilon_{pq}^{0(ij)} - \varepsilon_{pq}^{*(ij)} \right) E_{pqrs} \left( \varepsilon_{rs}^{0(kl)} - \varepsilon_{rs}^{*(kl)} \right) \boldsymbol{\varphi}(\mathbf{x})^{T} \delta(\Phi) d\Omega \right\} \frac{d\boldsymbol{\alpha}(t)}{dt} \tag{20}$$

Comparing Eq. (20) with the derivative of elasticity parameter $E_{ijkl}^{H}$ with respect to time variable $t$ by using the chain rule, the first-order derivatives of homogenized elasticity tensor with respect to the expansion coefficients can be given by:

$$\frac{\partial E_{ijkl}^{H}}{\partial \boldsymbol{\alpha}} = \frac{1}{|Y|} \int_{D} \left( \varepsilon_{pq}^{0(ij)} - \varepsilon_{pq}^{*(ij)} \right)^{T} E_{pqrs} \left( \varepsilon_{rs}^{0(kl)} - \varepsilon_{rs}^{*(kl)} \right) \boldsymbol{\varphi}(\mathbf{x})^{T} \delta(\Phi) d\Omega \tag{21}$$

Similarly, the first-order derivatives of volume constraint with respect to the design variables are derived as follows:

$$\frac{\partial V}{\partial \boldsymbol{\alpha}} = \frac{1}{|Y|} \int_{D} \boldsymbol{\varphi}(\mathbf{x})^{T} \delta(\Phi) d\Omega \tag{22}$$

## 5 Numerical implementations

After obtaining the design sensitivities, the OC method [Zhou and Rozvany (1991)] is employed to update the design variables. We introduce the regularized design variables $\bar{\alpha}_j$ for the actual design variables because it is difficult to specify the lower and upper bounds during the optimization process. The OC-based heuristic scheme is stated as:

**Step 1**. Defining the Lagrange function for the optimization formulation by the Lagrange multipliers $\Lambda$, $\lambda_1$ and $\lambda_2$, given as:

$$L = J + \Lambda G + \lambda_1 \left( \bar{\alpha}_j^{L} - \bar{\alpha}_j \right) + \lambda_2 \left( \bar{\alpha}_j - \bar{\alpha}_j^{U} \right) \tag{23}$$

**Step 2**. Calculating the regularized design variables:

$$\bar{\alpha}_j^{k} = \frac{\alpha_j^{k} - \alpha_{\min}^{k}}{\alpha_{\max}^{k} - \alpha_j^{k}} \tag{24}$$

where $k$ denotes the current iteration step starting from 1. $\alpha_{\min}^{k}$ and $\alpha_{\max}^{k}$ are defined as:

$$\begin{cases} \alpha_{\min}^k = 2 \times \min\left(\alpha_j^k\right) \\ \alpha_{\max}^k = 2 \times \max\left(\alpha_j^k\right) \end{cases} \tag{25}$$

**Step 3**. Based on the Kuhn-Tucker conditions [Zhou and Rozvany (1991)], the regularized design variables can be iteratively updated by using a heuristic scheme, given by:

$$\bar{\alpha}_j^{k+1} = \begin{cases} \max\{(\bar{\alpha}_j^k - m), \bar{\alpha}^L\}, & \text{if } \left(D_j^k\right)^\zeta \bar{\alpha}_j^k \leq \max\{(\bar{\alpha}_j^k - m), \bar{\alpha}^L\} \\ \left(D_j^k\right)^\zeta \bar{\alpha}_j^k, & \text{if } \begin{cases} \max\{(\bar{\alpha}_j^k - m), \bar{\alpha}^L\} < \left(D_j^k\right)^\zeta \bar{\alpha}_j^k < \\ \min\{(\bar{\alpha}_j^k + m), \bar{\alpha}^U\} \end{cases} \\ \min\{(\bar{\alpha}_j^k + m), \bar{\alpha}^U\}, & \text{if } \min\{(\bar{\alpha}_j^k + m), \bar{\alpha}^U\} \leq \left(D_j^k\right)^\zeta \bar{\alpha}_j^k \end{cases} \tag{26}$$

where $m$ and $\zeta$ are the move limit and the damping factor, respectively. The term $D_j^k$ stands for the updating factor based on the sensitivities of the objective and constraint functions, given as:

$$D_j^k = \frac{\partial E_{ijkl}^H}{\partial \alpha_j^k} \bigg/ \left(\max\left(\mu, \Lambda^k \frac{\partial V}{\partial \alpha_j^k}\right)\right) \tag{27}$$

where the Lagrange multiplier $\Lambda$ can be updated by a bi-sectioning algorithm (Sigmund and Maute 2013).

**Step 4**. Computing the actual design variables:

$$\alpha_j^k = \bar{\alpha}_j^k \times \left(\alpha_{\max}^k - \alpha_{\min}^k\right) + \alpha_{\min}^k \tag{28}$$

**Step 5**. Repeating Step 2 to Step 4 until the convergent criterion is satisfied.

## 6 Numerical examples

In this section, several numerical examples corresponded to the specific mechanical properties are provided to display the effectiveness of the proposed optimization formulation. In all examples, the Young's modulus for the solid material and the void phase are $E_0^s = 1$ and $E_0^v = 0.001$ respectively, and the Poisson's ratio is $\mu = 0.3$. For numerical simplicity, the dimensions of material cells are defined as $1 \times 1$, which is discretized by a mesh of $100 \times 100$. In the OC algorithm, the moving limit is $m = 0.01$ and the damping factor is set to $\zeta = 0.3$. A simple but efficient "Ersatz Material" model is employed to calculate the strains of the moving boundary cut by the meshes [Allaire et al. (2004)]. The optimization will be terminated when the difference of the objective between two successive steps is less than 10⁻³ or the maximum 100 iterative steps are reached. For numerical simplicity, the grids associated with the CSRBFs are supposed to be identical with the FE meshes.

### *6.1 Micro-structured materials with maximum bulk modulus*

The main intention of this example is to obtain the micro-structured materials with the maximum bulk modulus. According to the definition of the bulk modulus $K$, the objective function is stated as:

$$J = 4K = \sum_{i,j=1}^{2} E_{iijj}^{H} \tag{29}$$

In order to discuss the influence of the initial design of the material cell on the optimized solution, four kinds of initial designs are displayed in **Fig. 3**. The initial design 1 in case 1 is fully filled with solid materials in the design domain. 3×3 holes are uniformly distributed in the interior of the initial design 2 in case 2, while the voids are only arranged in the structural boundaries of the initial design 3 in case 3. The initial design 4 in case 4 can be regarded as the combination of the initial designs 2 and 3. The maximum volume fractions in four cases are all set to be 0.3. Meanwhile, a mesh of 100×100 finite elements is utilized to uniformly discretize the material cell. As an illustration, the initial design 3 is discretized into 16×16 finite elements, shown in **Fig. 4** and an exemplified finite element are shown in the right. The stiffness of the element cut by the moving boundary will be approximately calculated by the ersatz material model [Allaire et al. (2004)].

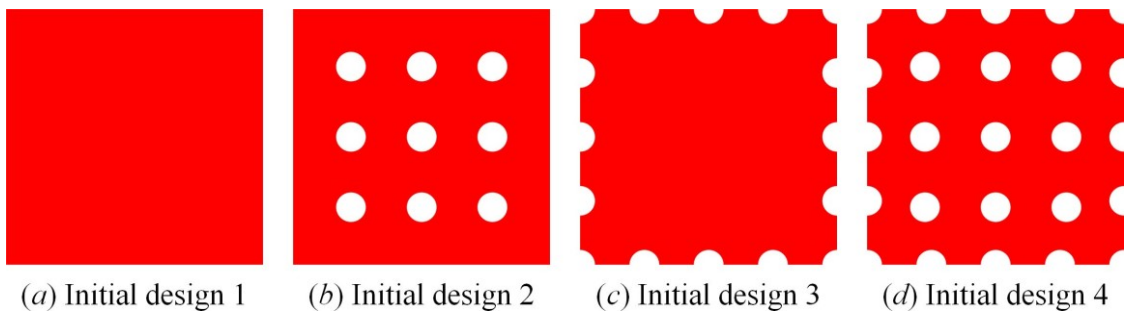

(*a*) Initial design 1    (*b*) Initial design 2    (*c*) Initial design 3    (*d*) Initial design 4

**Fig. 3**. Initial designs of material cells

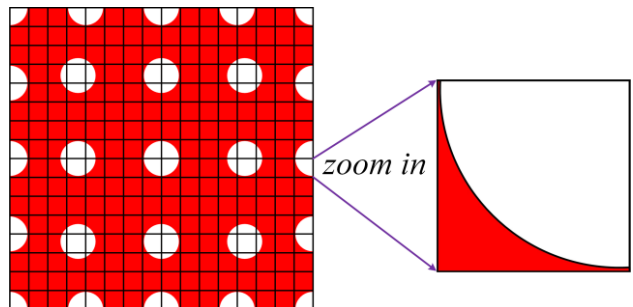

**Fig. 4** finite element mesh

The optimized results in four cases listed in Table. 1, including the optimized topologies of material cells, the level set surfaces, 3×3 repetitive material cells, the corresponding homogenized elastic tensors and the optimized bulk moduli. It can be clearly seen that the optimal topologies of material cells with maximum bulk modulus are common with the previous works [Cadman et al. (2013); Radman et al. (2013); Xia and Breitkopf (2015)], which illustrates the effectiveness of the proposed optimization formulation in the current work. Additionally, compared with the previous topologies of material cells, the obtained material cells in this paper are featured with smooth boundaries, clear interfaces between solids and voids. The geometrical features of the optimized results are beneficial to the latter manufacturing phase, due to that we can easily extract the structural geometries based on the optimized structure. The main reason is that the implicit boundary formulation in the PLSM to present structural boundaries, where the structural boundaries are implicitly embedded in the zero level-set of the level set function. The corresponding level set surfaces are displayed in forth column of Table. 1. Meanwhile, it is well-known that the Hashin-Shtrikman (HS) bounds [Hashin et al. (1963)] can be applied to evaluate the lower and upper properties for quasi-homogeneous and quasi-isotropic composites, given as:

$$K_{HS}^{up} = \frac{VK_S G_S}{(1-V)K_S + G_S} \tag{30}$$

where $V$ is the volume fraction of the material cell. $K_S$ and $G_S$ are the bulk modulus of the solid materials, respectively. $K_{HS}^{up}$ is the corresponding upper bound for the bulk modulus of the material cell with the volume fraction $V$. Hence, the upper bound $K_{HS}^{up}$ is equal to 0.095. It can be easily seen that the final optimized bulk moduli in four cases are very close to the upper bound.

In the third column of Table. 1, the configurations of optimized material cells with maximum bulk modulus are completely different under different initializations in four cases, and the corresponding homogenized elastic tensors are also varied with respect to the initial designs, shown in the sixth column of Table. 1. It is reasonable that there are many local optimums in terms of the design of micro-structured materials, and it is practically difficult to obtain a global optimal [Gao et al. (2018); Osanov and Guest (2016); Sigmund (1994); Xia and Breitkopf (2015)]. Moreover, it is known that many different topologies of material cells still correspond to the same homogenized elastic tensor. Although the optimized configurations and homogenized elastic tensor of material cells in four cases are different, the final bulk moduli in four cases are roughly equal, as shown in the final column of Table. 1. Hence, we confirm that the proposed optimization formulation is suitable for the design of micro-structured materials to attain maximum bulk modulus.

Table. 1 Optimized results in four cases

| Case | $V_{max}$ | Material cell | Level set surface | 3×3 material cells | $\mathbf{E}^H$ | $K$ |
|------|-----------|---------------|-------------------|--------------------|-----------------|-----|
| 1 | 0.3 | 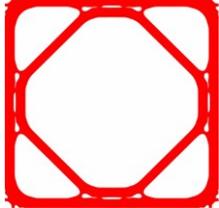 | 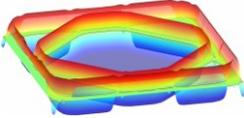 | 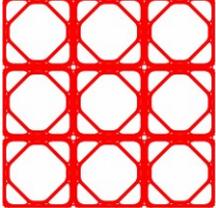 | $\begin{bmatrix} 0.142 & 0.046 & 0 \\ 0.046 & 0.142 & 0 \\ 0 & 0 & 0.026 \end{bmatrix}$ | 0.094 |
| 2 | 0.3 | 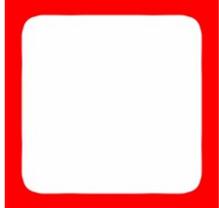 | 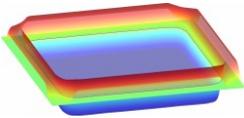 | 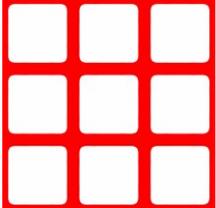 | $\begin{bmatrix} 0.173 & 0.014 & 0 \\ 0.014 & 0.173 & 0 \\ 0 & 0 & 0.004 \end{bmatrix}$ | 0.093 |
| 3 | 0.3 | 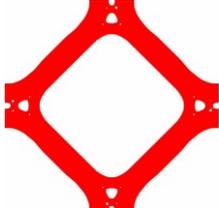 | 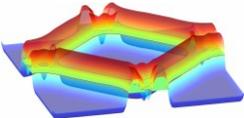 | 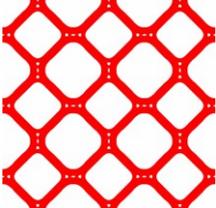 | $\begin{bmatrix} 0.099 & 0.088 & 0 \\ 0.088 & 0.099 & 0 \\ 0 & 0 & 0.046 \end{bmatrix}$ | 0.094 |
| 4 | 0.3 | 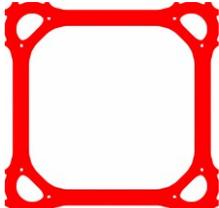 | 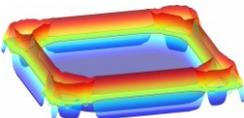 | 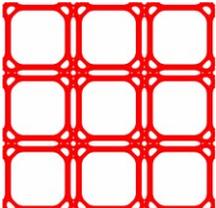 | $\begin{bmatrix} 0.152 & 0.037 & 0 \\ 0.037 & 0.152 & 0 \\ 0 & 0 & 0.003 \end{bmatrix}$ | 0.094 |

The iterative histories of the objective, volume fraction and the configurations in four cases are displayed in **Fig. 5**. It can be found that the objective changes remarkably at initial 10 steps because the violation of the volume constraint, and then it starts to increase stably when the volume function keeps constant. All the curves have a smooth, fast and stable convergence, which illustrates the high optimization efficiency of the proposed method. This mainly results from that the PLSM mitigates some unfavorable numerical schemes of the LSM aroused from directly solving the H-J PDEs [Li et al. (2016); Luo et al. (2008)]. Meanwhile, it is easily seen that the total iterative steps are different with the initial designs of material cells, while the optimized objectives are roughly equal. Hence, we can confirm that the initial design significantly affects the optimization efficiency of micro-structured materials, while has a negligible influence on the final solution. The distributed holes in the initial designs can trigger the topological changes in the initial steps, which can provide a search direction for the optimization design. As shown in **Fig. 5** (*a-d*), it is obviously found that new holes can be naturally created (**Fig. 5** (*a*)) and the existed holes can be merged gradually (**Fig. 5** (*c-d*)) to achieve the topological shape evolutions of material cells. Hence, we confirm that a suitable number of holes are required in the initial designs, in order to provide the initial design a good inhomogeneity that will benefit the convergence of the optimization.

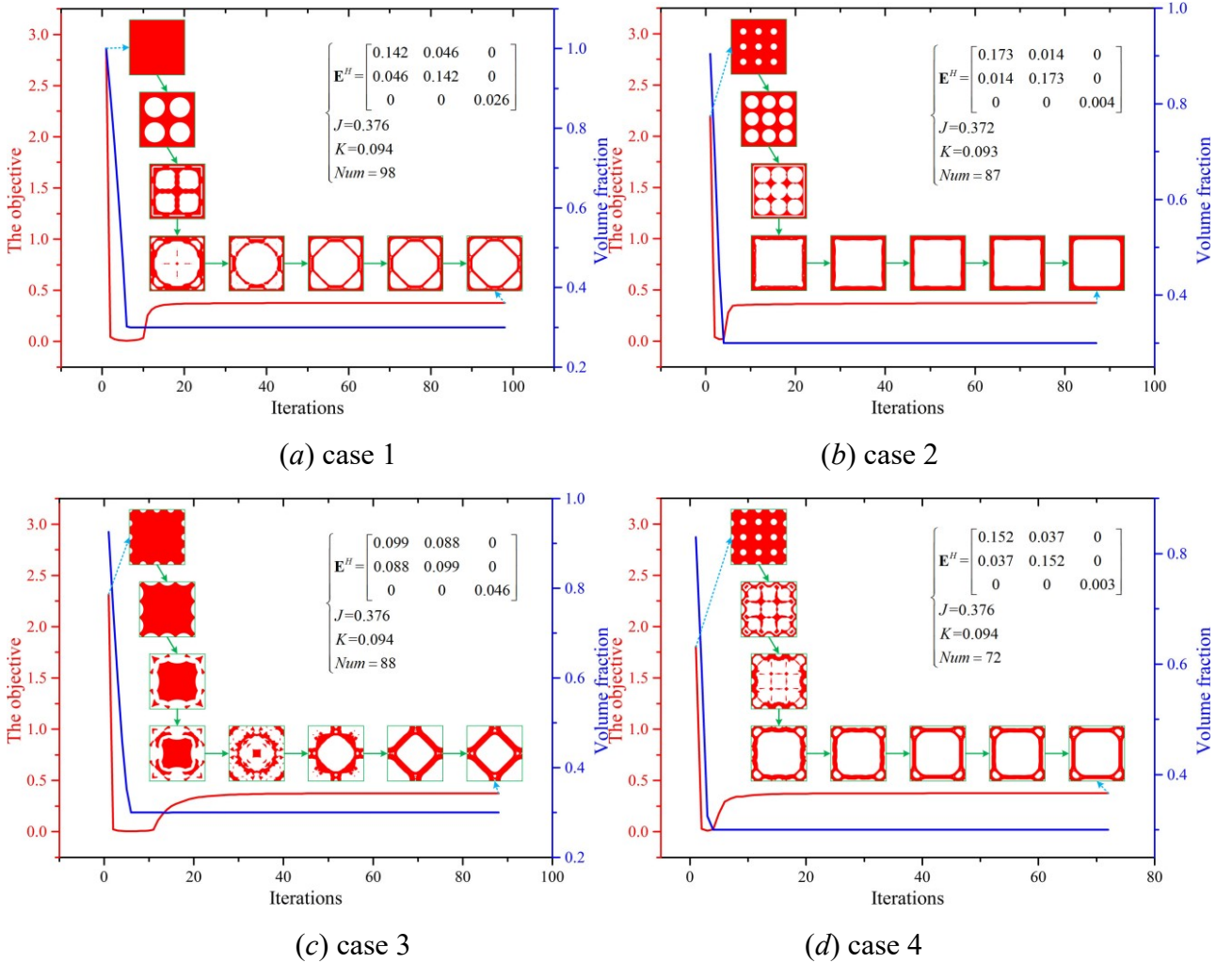

(*a*) case 1  (*b*) case 2

(*c*) case 3  (*d*) case 4

**Fig. 5**. Iterative curves of four cases

*6.2 Micro-structured materials with maximum shear modulus*

In this section, we apply the optimization formulation to gain the micro-structured materials with maximum shear modulus. The objective function is written as:

$$J = G = E^H_{1212} \tag{31}$$

To further show the effectiveness of the proposed optimization formulation, four cases are still addressed based on the definition of initial designs shown in **Fig. 3**. The design parameters in four cases are same as the previous example, while the volume fractions are set to be 0.4.

The optimized results in four cases are listed in Table. 2, which includes the optimized configurations of material cells, level set surfaces, 3×3 repetitive material cells, the corresponding homogenized effective elastic tensors and the optimized shear moduli. The material cells with the optimized shear moduli illustrates the effectiveness of proposed optimization formulation again. As shown in the third column of Table. 2, it can be seen that the initial designs have a remarkable effect on the optimized topologies, while the final numerical solution of the maximum shear moduli in four cases are nearly the same. The main reason is that the non-uniqueness of the topologies of material cells with respect to homogenized elastic tensor are existed [Liu et al. (2015)]. As shown in the third and fourth columns of Table. 2, all the final designs of material cells are featured with smooth structural boundaries and distinct material interfaces due to the use of the PLSM. As far as the proposed method is concerned, the capability of nucleating new holes within the design domain is also observed in this example. We conclude that the proposed formulation is flexible in handling the complex topological and shape changes, by deleting or generating holes as well as merging or splitting boundaries.

Table. 2 Optimized results in four cases

| Case | $V_{max}$ | Material cell | Level set surface | 3×3 material cells | $\mathbf{E}^H$ | $G$ |
|---|---|---|---|---|---|---|
| 1 | 0.4 | 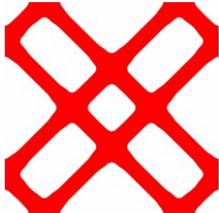 | 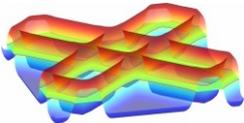 | 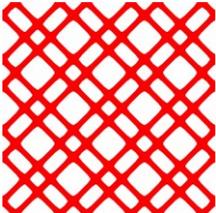 | $\begin{bmatrix} 0.144 & 0.126 & 0 \\ 0.126 & 0.144 & 0 \\ 0 & 0 & 0.108 \end{bmatrix}$ | 0.108 |
| 2 | 0.4 | 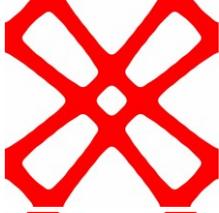 | 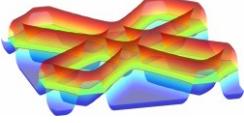 | 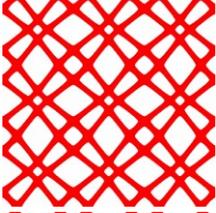 | $\begin{bmatrix} 0.142 & 0.123 & 0 \\ 0.123 & 0.142 & 0 \\ 0 & 0 & 0.109 \end{bmatrix}$ | 0.109 |
| 3 | 0.4 | 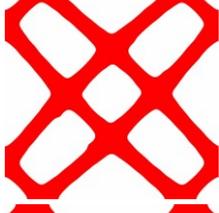 | 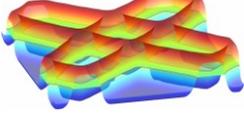 | 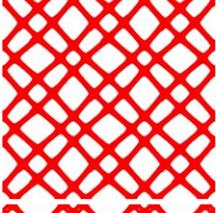 | $\begin{bmatrix} 0.145 & 0.123 & 0 \\ 0.123 & 0.145 & 0 \\ 0 & 0 & 0.108 \end{bmatrix}$ | 0.108 |
| 4 | 0.4 | 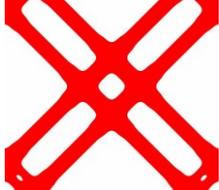 | 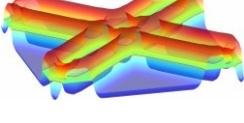 | 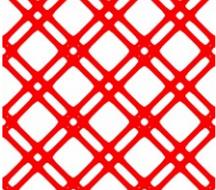 | $\begin{bmatrix} 0.142 & 0.129 & 0 \\ 0.129 & 0.142 & 0 \\ 0 & 0 & 0.11 \end{bmatrix}$ | 0.110 |

## 6.3 Micro-structured materials with multiple objective functions

In this section, we apply the formulation to optimize the micro-structured materials considering the maximum bulk modulus and the maximum shear modulus simultaneously. The objective function is defined as:

$$J = \omega_1 K + \omega_2 G \qquad (33)$$

where $\omega_1$ and $\omega_2$ are the corresponding weights for the bulk modulus and the shear modulus. In this example, they are defined to be 0.5 and 0.5. We also optimize the material cells based on four initial designs defined in **Fig. 3**. The design parameters in four cases are same as the first example. Table. 3 provides the optimized numerical results for four cases, including the optimized topologies of material cells, level set surfaces, 3×3 repetitive material cells, the homogenized effective elastic tensors and the optimized objective. Based on the optimized topologies in four cases, the effectiveness of the proposed optimization formulation can be further shown. Meanwhile, we can also observe that the initial designs remarkably affect the optimized topology, while the final objective in four cases are very close. The phenomena are similar to Sections 6.1 and 6.2.

Table. 3 Optimized results in four cases

| Case | $V_{max}$ | Material cell | Level set surface | 3×3 material cells | $\mathbf{E}^H$ | $K$ |
|---|---|---|---|---|---|---|
| 1 | 0.3 | 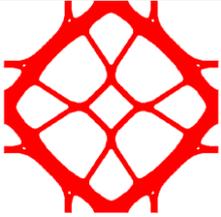 | 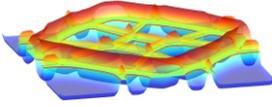 | 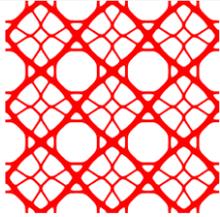 | $\begin{bmatrix} 0.103 & 0.084 & 0 \\ 0.084 & 0.103 & 0 \\ 0 & 0 & 0.016 \end{bmatrix}$ | 0.055 |
| 2 | 0.3 | 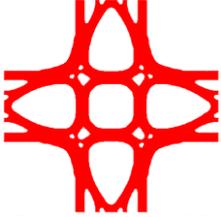 | 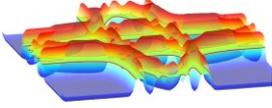 | 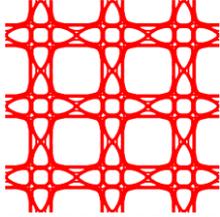 | $\begin{bmatrix} 0.146 & 0.04 & 0 \\ 0.04 & 0.146 & 0 \\ 0 & 0 & 0.016 \end{bmatrix}$ | 0.055 |
| 3 | 0.3 | 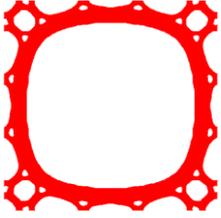 | 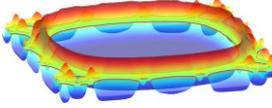 | 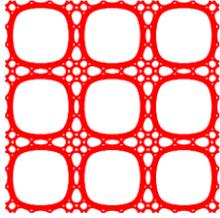 | $\begin{bmatrix} 0.129 & 0.059 & 0 \\ 0.059 & 0.129 & 0 \\ 0 & 0 & 0.013 \end{bmatrix}$ | 0.054 |
| 4 | 0.3 | 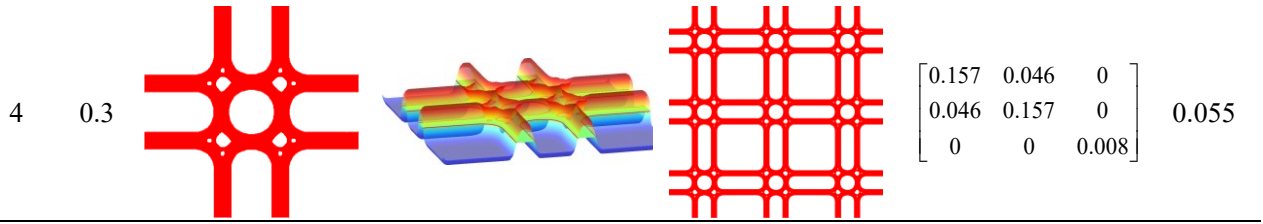 | | $\begin{bmatrix} 0.157 & 0.046 & 0 \\ 0.046 & 0.157 & 0 \\ 0 & 0 & 0.008 \end{bmatrix}$ | 0.055 |

### 6.4 Micro-structured materials with NPR

The main intention of this section is to gain micro-structured materials with the NPR. It is known that there exist two directions of the definition of Poisson's ratio in 2D micro-structured materials, i.e. $\upsilon_{12} = E^H_{1122}/E^H_{11}$ and $\upsilon_{21} = E^H_{1122}/E^H_{2222}$. For isotropic materials, the $\upsilon_{12}$ and $\upsilon_{21}$ are equal, it does not matter either $\upsilon_{12}$ or $\upsilon_{21}$ is used as an objection function to achieve the NPR. However, the equal relationship in two definitions of the Poisson's ratios is not existed in the orthotropic materials. Hence, a relaxed objective function is defined to obtain the NPRs in both the isotropic and orthotropic micro-structured materials, given by:

$$J = -E^H_{1212} + 0.03 \times \left( E^H_{1111} + E^H_{2222} \right) \tag{34}$$

It can be easily found that the defined objective function tends to maximize the horizontal and vertical stiffness moduli and minimize the $E^H_{1212}$ so that the objective is maximized and the NPR structure is obtained. Hence, it is suitable for us to define this objective to achieve the NPR structure.

In order to discuss the effectiveness of proposed optimization formulation on both the isotropic and orthotropic materials, we define six kinds of initial designs corresponded to six cases. As shown in Fig. 6, the isotropic and orthotropic cellular materials are defined by applying the geometrical symmetry to the material cells, where the orthotropy is achieved by applying two symmetries with regard to *x* and *y* directions to material cells in case 1 to 3 and the isotropy is obtained by applying the square and diagonal geometrical symmetries to material cells in cases 4-6. The maximum volume fractions in six cases are all defined to be 0.35.

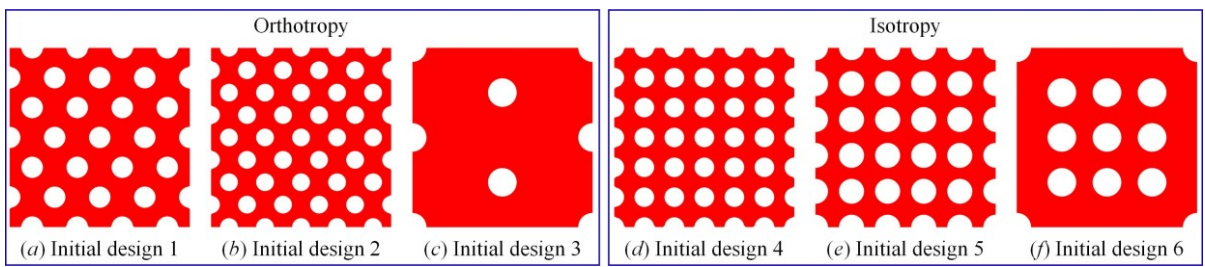

Fig. 6. Initial designs in six cases

The optimized results in six cases to attain the NPR are listed in Table. 4, including the optimized topologies of material cells, level set surfaces, 3×3 repetitive material cells, the corresponding homogenized elastic tensors and the optimized NPRs in two directions. The optimized material cells with the NPR in six cases display the effectiveness of proposed optimization formulation. Similar to the above examples, the optimized configurations of material cells with the NPR completely different in six cases, which significantly depend on the initial designs. As shown in the final column of Table. 4, it can be easily seen that the optimal 2D orthotropic micro-structured materials are featured with the NPR in two directions by the proposed objective function. Hence, the objective function can concurrent optimize the NPRs $\upsilon_{12}$ and $\upsilon_{21}$. Moreover, the Poisson's ratio for 2D orthotropic micro-structured materials can be lower than -1 (shown in cases 1, 2 and 3), while the 2D isotropic micro-structured materials has the lower bound of the Poisson's ratio (shown in case 4, 5 and 6). We also can see that a series of new and interesting micro-structured materials with the NPR are gained. As displayed in the third and fifth columns of Table. 4, the optimized topologies of material cells with the NPR have the smooth boundaries and are free of grayscales between solids and voids result from that the implicit representation of structural boundaries in the level set surfaces shown in the fourth column of Table. 4.

Table. 4 Optimized results in six cases

| Case | $V_{max}$ | Material cell | Level set surface | 3×3 material cells | $\mathbf{E}^H$ | $\upsilon$ |
|---|---|---|---|---|---|---|
| 1 | 0.35 | 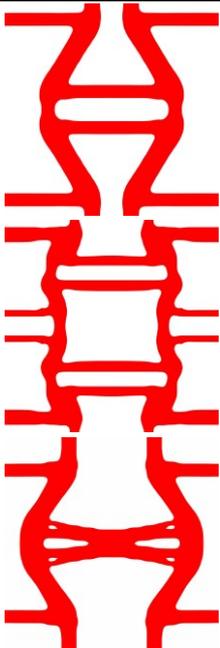 | 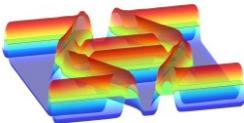 | 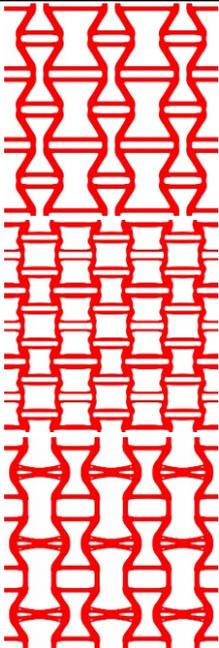 | $\begin{bmatrix} 0.037 & -0.048 & 0 \\ -0.048 & 0.098 & 0 \\ 0 & 0 & 0.002 \end{bmatrix}$ | $\upsilon_{12}=-1.3$ <br> $\upsilon_{21}=-0.49$ |
| 2 | 0.35 | 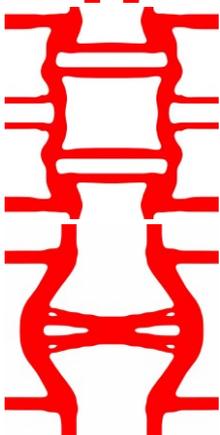 | 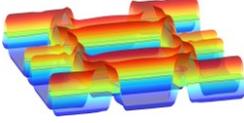 | 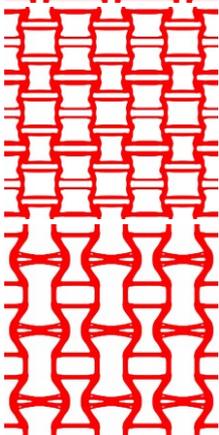 | $\begin{bmatrix} 0.103 & -0.053 & 0 \\ -0.053 & 0.0787 & 0 \\ 0 & 0 & 0.003 \end{bmatrix}$ | $\upsilon_{12}=-0.52$ <br> $\upsilon_{21}=-0.68$ |
| 3 | 0.35 | 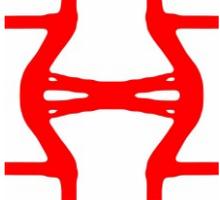 | 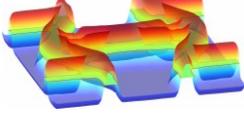 | 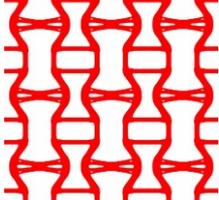 | $\begin{bmatrix} 0.43 & -0.048 & 0 \\ -0.048 & 0.102 & 0 \\ 0 & 0 & 0.004 \end{bmatrix}$ | $\upsilon_{12}=-1.12$ <br> $\upsilon_{21}=-0.48$ |

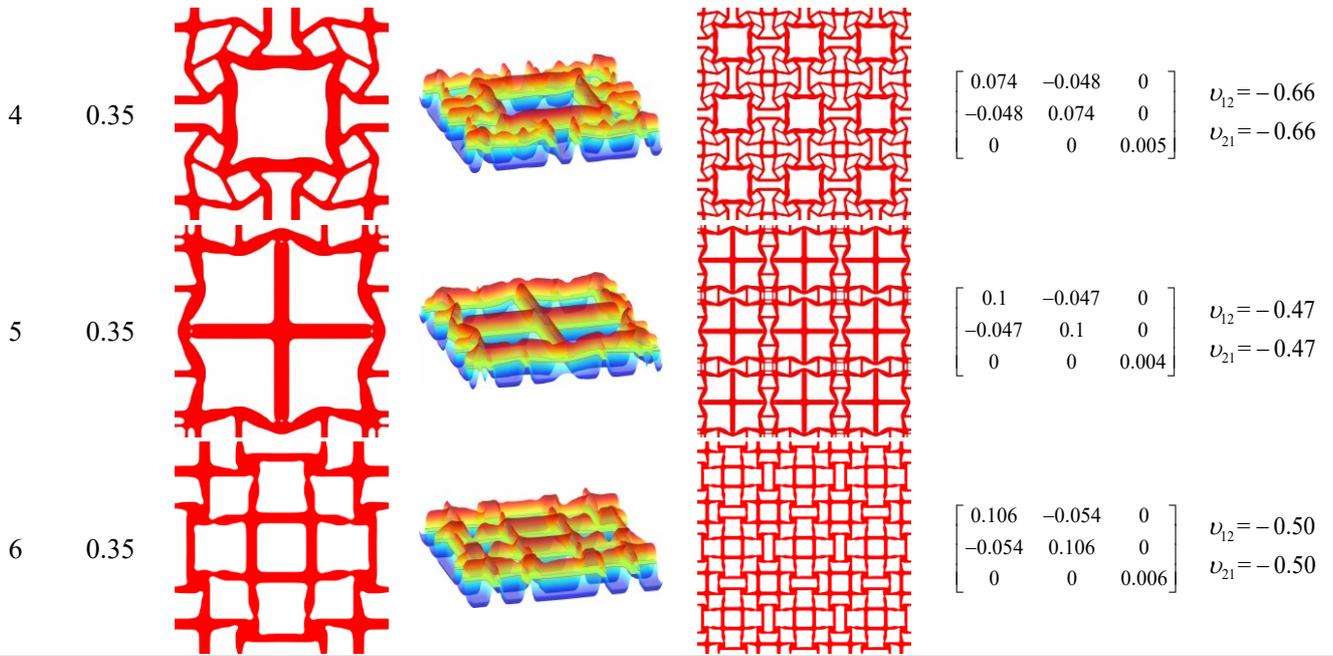

Meanwhile, the optimized iterations in case 3 and case 4 are displayed in Fig. 7. It can be easily seen that the trajectories of the convergent processes show that the objective and volume fraction converge to the optimal solutions rapidly with 100 steps, which shows the high optimization efficiency of the proposed optimization formulation. Moreover, the nucleation mechanism of new holes and the merging mechanism of the existed holes can be perfectly illustrated by the evolutions of topologies of material cells in Fig. 7.

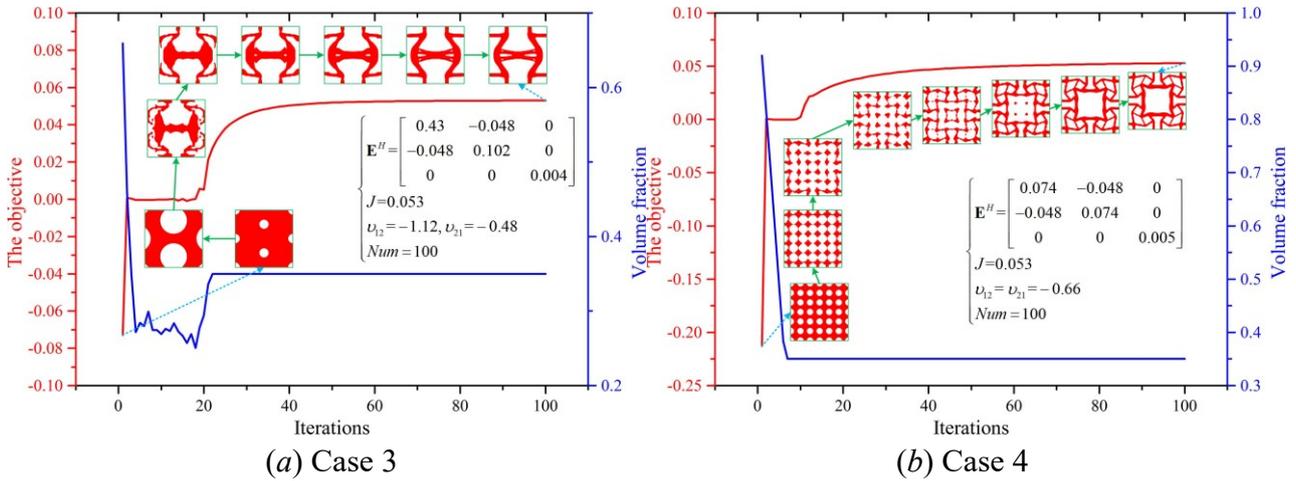

(a) Case 3    (b) Case 4

Fig. 7. Iterative curves of case 3 and case 4

In order to show the NPR characteristic of micro-structured materials, we input the above optimized result in case 4 into the engineering software ANASYS for simulating the deformation of the NPR structure. Firstly, we export the optimized NPR structure in a STL file format and input the ANSYS SpaceClaim Direct Modeler to reconstruct the NPR structure. We stretch the NPR structure along the depth direction (1 mm), and the stretched depth is much less than the length (100mm) and height (100mm) in order to maintain the plane stress condition to some extent. The NPR structure in ANSYS is shown in **Fig. 8**. We label the corresponding boundaries (the left boundaries A1, A2, A3 and A4; the opposite boundaries B1, B2, B3 and B4).

We fix three directional displacements of the boundaries A2 and A3. Meanwhile, both A1 and A4 only have the deformation in height direction compared with the fixed boundaries (A2 and A3). We impose a uniform displacement field (1mm) on the opposite boundaries B1, B2, B3 and B4 in length direction. The NPR structure is discretized by the body-fit finite elements in ANSYS, as shown in the right plot of **Fig. 8**. After solving, the displacement field of the NPR structure is displayed in **Fig. 9**. Meanwhile, the initial NPR structure is also plotted in the black line. Hence, it can be easily seen that the structure is featured with the NPR. That is, the structure expands along the height direction when stretched in the length direction.

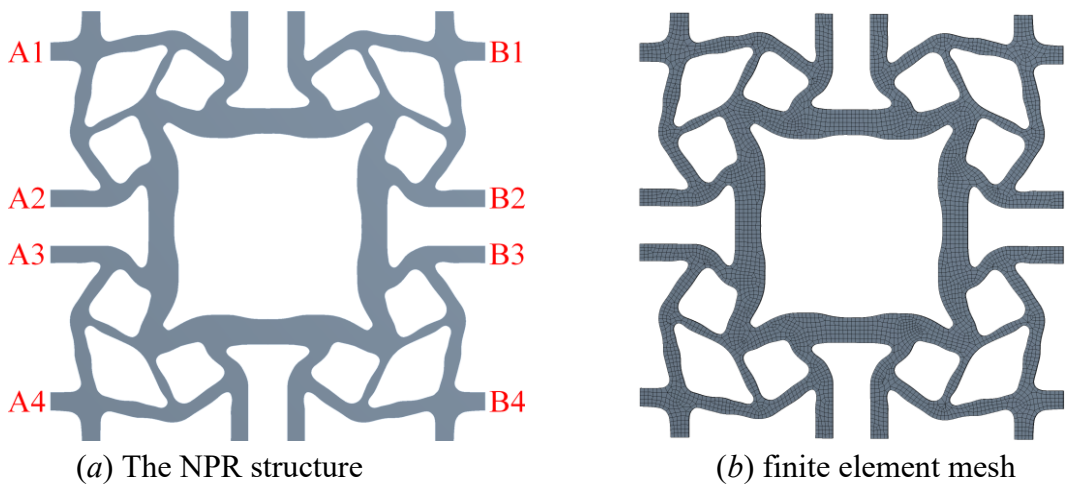

(*a*) The NPR structure　　　　　　　　　(*b*) finite element mesh
**Fig. 8.** The NPR structure in ANASYS

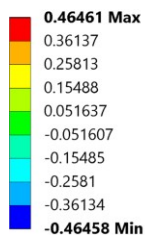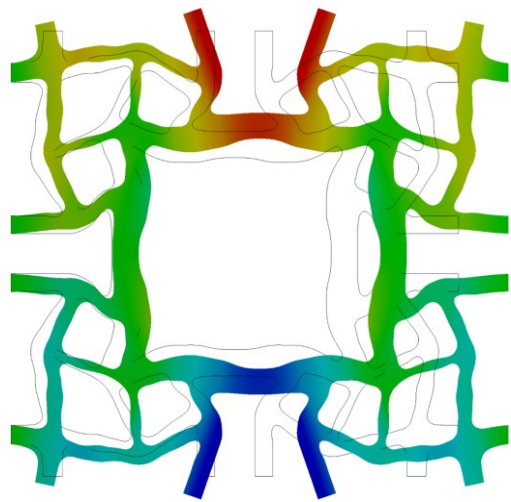

**Fig. 9**. The displacement field of the NPR structure along height direction

Based on the results in the simulation, the maximum displacement in the top or bottom boundaries is equal to 0.465. The displacements of the right boundaries in the length direction are all equal to 1mm. The Poisson's ratio is equal to -0.465. In the above computation, the NPR is equal to -0.66. Based on the careful considerations for this error between the optimized result and simulated result, four factors may contribute to the occurrence of the error to some extent: 1) In the above optimization, the simple ersatz material model is an approximate method to evaluate the element stiffness, which extensively influence the numerical precision of the numerical results. 2) When the element is cut by the structural boundary, it is difficult to exactly capture the structural geometry and calculate the element volume fraction accurately. 3) When input the STL file into the ANASYS to obtain the NPR structure, there are many fine facets in the inputted structure. The reverse engineering in the ANSYS SpaceClaim to rebuild the NPR structure result in the difference to some extent compared with the initial geometry. 4) The quality of the finite element mesh in ANASYS also significantly affects the final result.

However, it should be noted that the improvement of computational precision is not within the scope of the current work. The key contribution is the development of an effective optimization formulation for the micro-structured materials with the specific mechanical properties.

## 7. Conclusions

This paper performs the topology optimization design of micro-structured materials featured with the specific mechanical properties using the PLSM combined with the EBHM. The EBHM is applied to evaluate material effective properties working as the objective based on the topology of the material cell. The PLSM is used to optimize structural topology to find the expected effective elastic properties. Four numerical examples are given to present the effectiveness of proposed optimization formulation, including the maximum bulk modulus, maximum shear modulus and their combination, as well as the negative Poisson's ratio.

A series of new and interesting micro-structured materials with specific mechanical properties are achieved. The solutions show that the structural topologies are featured with smooth boundaries, clear interfaces. As far as the numerical implementations and the optimized results is concerned, several favourable features of the proposed optimization design formulation are involved such that it can be regarded as a general approach especially serve for the topological shape design of micro-structured materials.

**Acknowledgments**

This work was supported by the National Basic Scientific Research Program of China [grant number JCKY2016110C012], and National Natural Science Foundation of China [grant number 51705166] and also Australian Research Council (ARC) - Discovery Projects (160102491).

**References**

Allaire, G., Jouve, F., Toader, A.M. [2004] "Structural optimization using sensitivity analysis and a level-set method," J. Comput. Phys. **194**, 363–393.

Ashby, M.F., Evans, T., Fleck, N.A., Hutchinson, J., Wadley, H., Gibson, L. [2000] "Metal foams: a design guide," Elsevier.

Bendsøe, M.P., Kikuchi, N. [1988] "Generating optimal topologies in structural design using a homogenization method," Comput. Methods Appl. Mech. Eng. **71**, 197–224.

Bendsoe, M.P., Sigmund, O. [1999] "Material interpolation schemes in topology optimization," Arch. Appl. Mech. **69**, 635–654.

Berger, J., Wadley, H., McMeeking, R. [2017] "Mechanical metamaterials at the theoretical limit of isotropic elastic stiffness," Nature **543**, 533-537.

Buhmann, M.D. [2003] "Radial basis functions: theory and implementations," vol. 12. Cambridge university press.

Cadman, J.E., Zhou, S., Chen, Y., Li, Q. [2013] "On design of multi-functional microstructural materials," J. Mater. Sci. **48**, 51-66.

Challis, V.J., Guest, J.K., Grotowski, J.F., Roberts, A.P. [2012] "Computationally generated cross-property bounds for stiffness and fluid permeability using topology optimization," Int. J. Solids. Struct. **49**, 3397-3408.

Christensen, R.M. [2000] "Mechanics of cellular and other low-density materials," Int. J. Solids. Struct. **37**, 93-104.

Chu S, Gao L, Xiao M. [2018] "An efficient topology optimization method for structures with uniform stress," Int. J. Comput. Methods. **15(1),** 1850073.

Da, D., Chen, J., Cui, X., Li, G. [2017] "Design of materials using hybrid cellular automata," Struct. Multidiscip. Optim. **56**, 131-137.

Dunning, P.D., Alicia Kim, H. [2013] "A new hole insertion method for level set based structural topology optimization," Int. J. Numer. Methods. Eng. **93**, 118-134.

Gao T, Qiu L, Zhang W. [2017] "Topology optimization of continuum structures subjected to the variance constraint of reaction forces," Struct. Multidiscip. Optim. **56**(4), 755-765.

Groen J P, Sigmund O. [2018] "Homogenization‐based topology optimization for high‐resolution manufacturable microstructures," Int. J. Numer. Methods. Eng. **113**(8), 1148-1163.

Gao, J., Li, H., Gao, L., Xiao, M. [2018] "Topological shape optimization of 3D micro-structured materials using energy-based homogenization method," Adv. Eng. Softw. **116**, 89-102.

Gibson, L.J., Ashby, M.F. [1999] "Cellular solids: structure and properties," Cambridge university press.

Guedes, J., Kikuchi, N. [1990] "Preprocessing and postprocessing for materials based on the homogenization method with adaptive finite element methods," Comput. Methods Appl. Mech. Eng. **83**, 143-198.

Guest, J.K., Prévost, J.H. [2006] "Optimizing multifunctional materials: design of microstructures for maximized stiffness and fluid permeability," Int. J. Solids. Struct. **43**, 7028-7047.

Guo, X., Zhang, W., Zhong, W. [2014] "Doing Topology Optimization Explicitly and Geometrically—A New Moving Morphable Components Based Framework," J. Appl. Mech. **81**, 081009.

Rodrigues H, Guedes JM, Bendsoe MP [2002] "Hierarchical optimization of material and structure," Struct. Multidiscip. Optim. **24**, 1-10.

Hassani, B., Hinton, E. [1998a] "A review of homogenization and topology optimization I - homogenization theory for media with periodic structure," Comput. Struct. **69**, 707-717.

Hassani, B., Hinton, E. [1998b] "A review of homogenization and topology optimization II - analytical and numerical solution of homogenization equations," Comput. Struct. **69**, 719-738.

Hashin Z, Shtrikman S. [1963] "A variational approach to the theory of the elastic behaviour of multiphase materials," J. Mech. Phys. Solids, **11**(2), 127-140.

Huang X, Radman A, Xie Y [2011] "Topological design of microstructures of cellular materials for maximum bulk or shear modulus" Comp Mater Sci. **50**:1861-1870.

Kang Z, Wang Y. [2011] "Structural topology optimization based on non-local Shepard interpolation of density field," Comput. Methods Appl. Mech. Eng. **200** (49-52), 3515-3525.

Kang, Z., Wang, Y., Wang, Y. [2016] "Structural topology optimization with minimum distance control of multiphase embedded components by level set method," Comput. Methods Appl. Mech. Eng. **306**, 299-318.

Li E, Zhang Z, Chang C C, et al. [2015] "Numerical homogenization for incompressible materials using selective smoothed finite element method". Compos. Struct. **123**, 216-232.

Li E, Zhang Z, Chang C C, et al. [2016] "A new homogenization formulation for multifunctional composites". Int. J. Comput. Methods. **13(02)**, 1640002.

Li, H., Li, P., Gao, L., Zhang, L., Wu, T. [2015] "A level set method for topological shape optimization of 3D structures with extrusion constraints," Comput. Methods Appl. Mech. Eng. **283**, 615-635.

Li H, Luo Z, Gao L, Walker P [2018] "Topology optimization for functionally graded cellular composites with metamaterials by level sets," Comput. Methods Appl. Mech. Eng. **328**, 340-364.

Li, H., Luo, Z., Zhang, N., Gao, L., Brown, T. [2016] "Integrated design of cellular composites using a level-set topology optimization method," Comput. Methods Appl. Mech. Eng. **309**, 453-475.

Li, Z., Shi, T., Xia, Q. [2017] "Eliminate localized eigenmodes in level set based topology optimization for the maximization of the first eigenfrequency of vibration," Adv. Eng. Softw. **107** [Supplement C], 59-70.

Liu, R., Kumar, A., Chen, Z., Agrawal, A., Sundararaghavan, V., Choudhary, A. [2015] "A predictive machine learning approach for microstructure optimization and materials design," Sci. Rep. **5**.

Long, K., Du, X., Xu, S., Xie, Y.M. [2016] "Maximizing the effective Young's modulus of a composite material by exploiting the Poisson effect," Compos. Struct. **153**, 593-600.

Luo, Z., Tong, L.Y., Kang, Z. [2009] "A level set method for structural shape and topology optimization using radial basis functions," Comput. Struct. **87**, 425-434.

Luo, Z., Wang, M.Y., Wang, S., Wei, P. [2008] "A level set-based parameterization method for structural shape and topology optimization," Int. J. Numer. Methods. Eng. **76**, 1-26.

Ma, Z.D., Kikuchi, N., Hagiwara, I. [1993] "Structural topology and shape optimization for a frequency response problem," Comput. Mech. **13**, 157-174.

McClanahan D R, Liu G R, Turner M G, et al. [2018] "Topology optimization of the interior structure of blades with an outer surface determined through aerodynamic design". Int. J. Comput. Methods. **15(3)**, 1840027.

Michel, J.-C., Moulinec, H., Suquet, P. [1999] "Effective properties of composite materials with periodic microstructure: a computational approach," Comput. Methods Appl. Mech. Eng. **172**, 109-143.

Osanov, M., Guest, J.K [2016] "Topology optimization for architected materials design," Ann. Rev. Mater. Res. **46**, 211-233.

Osher, S., Fedkiw, R. [2006] "Level set methods and dynamic implicit surfaces," vol. 153. Springer Science & Business Media.

Osher, S., Sethian, J.A. [1988] "Fronts propagating with curvature-dependent speed: algorithms based on Hamilton-Jacobi formulations" J. Comput. Phys. **79**, 12-49.

Radman, A., Huang, X., Xie, Y. [2013] "Topological optimization for the design of microstructures of isotropic cellular materials," Eng. Optimiz. **45**, 1331-1348.

Radman A, Huang X, Xie YM [2012] "Topology optimization of functionally graded cellular materials," J. Mater. Sci. **48,** 1503-1510.

Rozvany, G., Bendsøe, M., Kirsch, U. [1996] "Addendum-Layout optimization of structures," Appl. Mech. Rev. **49**, 54-54.

Sethian, J.A., Wiegmann, A. [2000] "Structural boundary design via level set and immersed interface methods," J. Comput. Phys. **163**, 489-528.

Sigmund, O. [1994] "Materials with prescribed constitutive parameters: an inverse homogenization problem," Int. J. Solids. Struct. **31**, 2313-2329.

Sigmund, O., Maute, K. [2013] "Topology optimization approaches: a review," Struct. Multidiscip. Optim. **48**, 1031-1055.

Svanberg, K. [1987] "The method of moving asymptotes—a new method for structural optimization," Int. J. Numer. Methods. Eng. **24**, 359-373.

Torquato, S., Hyun, S., Donev, A. [2003] "Optimal design of manufacturable three-dimensional composites with multifunctional characteristics," J. Appl. Phys. **94**, 5748.

Wang, M.Y., Wang, X. [2004] ""Color" level sets: a multi-phase method for structural topology optimization with multiple materials," Comput. Methods Appl. Mech. Eng. **193**, 469-496.

Wang, M.Y., Wang, X., Guo, D. [2003] "A level set method for structural topology optimization," Comput. Methods Appl. Mech. Eng. **192**, 227-246.

Wang, S., Wang, M.Y. [2006] "Radial basis functions and level set method for structural topology optimization," Int. J. Numer. Methods. Eng. **65**, 2060-2090.

Wang, X., Mei, Y., Wang, M.Y. [2005] "Level-set method for design of multi-phase elastic and thermoelastic materials," Int. J. Mech. Mater. Des. **1**, 213-239.

Wang Y, Chen F, Wang MY [2017] "Concurrent design with connectable graded microstructures," Comput. Methods Appl. Mech. Eng. **317**, 84-101.

Wang Y, Zhang L, Daynes S, et al. [2018] "Design of graded lattice structure with optimized mesostructures for additive manufacturing," Mater. Des. **142**, 114-123.

Wang Y, Gao J, Kang Z. [2018] "Level set-based topology optimization with overhang constraint: Towards support-free additive manufacturing," Comput. Methods Appl. Mech. Eng. **339**, 591-614.

Wang C, Zhu JH, Zhang WH, Li SY, Kong J [2018] "Concurrent topology optimization design of structures and non-uniform parameterized lattice microstructures," Struct. Multidiscip. Optim. DOI: 10.1007/s00158-018-2009-0

Wang, Y., Luo, Z., Kang, Z., Zhang, N. [2015] "A multi-material level set-based topology and shape optimization method," Comput. Methods Appl. Mech. Eng. **283**, 1570-1586.

Wang Y, Luo Z, Zhang N, Kang Z [2014] "Topological shape optimization of microstructural metamaterials using a level set method," Comp Mater Sci. **87**, 178-186.

Wendland, H. [1995] "Piecewise polynomial, positive definite and compactly supported radial functions of minimal degree," Adv. Comput. Math. **4**, 389-396.

Wu, J., Luo, Z., Li, H., & Zhang, N. [2017] "Level-set topology optimization for mechanical metamaterials under hybrid uncertainties," Comput. Methods Appl. Mech. Eng. **319**, 414-441.

Xia L, Breitkopf P [2015] "Multiscale structural topology optimization with an approximate constitutive model for local material microstructure," Comput. Methods Appl. Mech. Eng. **286**,147-167

Xia, L., Breitkopf, P. [2015] "Design of materials using topology optimization and energy-based homogenization approach in Matlab," Struct. Multidiscip. Optim. **52**, 1229-1241.

Xia, L., Xia, Q., Huang, X., Xie, Y.M. [2016] "Bi-directional evolutionary structural optimization on advanced structures and materials: a comprehensive review," Arch. Computat. Methods. Eng. **25**(2), 437-478.

Xia, Q., Wang, M.Y., Wang, S., Chen, S. [2006] "Semi-Lagrange method for level-set-based structural topology and shape optimization," Struct. Multidiscip. Optim. **31**, 419-429.

Xiao Mi, Chu Sheng, Gao Liang, Li Hao. [2018] "A hybrid method for density-related topology optimization," Int. J. Comput. Methods. **15(3).** 1850116.

Xie, Y.M., Steven, G.P. [1993] "A simple evolutionary procedure for structural optimization," Comput. Struct. **49**, 885-896.

Zhang W, Zhou L. [2018] "Topology optimization of self-supporting structures with polygon features for additive manufacturing," Comput. Methods Appl. Mech. Eng. **334**, 56-78.

Zheng, J., Yang, X., Long, S. [2014] "Topology optimization with geometrically non-linear based on the element free Galerkin method," Int. J. Mech. Mater. Des. **11**, 231-241.

Zhou, M., Rozvany, G. [1991] "The COC algorithm, Part II: Topological, geometrical and generalized shape optimization," Comput. Methods Appl. Mech. Eng. **89**, 309-336.